\documentclass[a4paper,12pt, twocolumn]{article}
\usepackage[utf8]{inputenc}
\usepackage{geometry}
\usepackage{graphicx}
\usepackage{amsmath}
\usepackage{caption}
\usepackage{titlesec}
\usepackage[dvipsnames]{xcolor}     
\usepackage{lipsum}     
\usepackage{times}      
\usepackage{ragged2e}   
\usepackage{float}
\usepackage{siunitx}
\usepackage{titling} 

\usepackage[backend=biber, style=numeric, sorting=none,giveninits=true]{biblatex}
\justifying                         

\titleformat{\section}{\normalfont\normalsize\bfseries\MakeUppercase}{\thesection.}{1em}{}
\titleformat{\subsection}{\normalfont\normalsize\bfseries}{\thesubsection.}{1em}{}
\titleformat{\subsubsection}{\normalfont\normalsize\bfseries}{\thesubsubsection.}{1em}{}

{
\centering

\title{\fontsize{16pt}{19pt}\selectfont
\textbf{Robust and Reconfigurable On-Board Data Handling} \\
\textbf{Subsystem for Present and Future Brazilian} \\
\textbf{CubeSat Missions}
\vspace{-2em}}

\author{
    \fontsize{12pt}{14pt}\selectfont
    \textbf{Victor O. Costa (1), Mauren D'Ávila (1), Douglas Arena (1), Vinicius Schreiner (1),}\\[-4pt]
    \fontsize{12pt}{14pt}\selectfont
    \textbf{Renan Menezes (1), Cleber Hoffmann (1), Edson Pereira (1), Lidia Shibuya Sato (1),}\\[-4pt]
    \fontsize{12pt}{14pt}\selectfont\textbf{Felipe Tavares (1), Luis Loures (1), Fernanda L. Kastensmidt (2)}\\
    \\[-4pt]
    \fontsize{12pt}{14pt}\selectfont
    (1) Instituto Tecnológico de Aeronáutica - Centro Espacial ITA, Brasil\\[-4pt]
    \fontsize{12pt}{14pt}\selectfont
    (2) Universidade Federal do Rio Grande do Sul - PGMICRO, Brasil \\
    \vspace{-2.5em}}
}
\date{}

\begin{document}

\maketitle

\begin{abstract}
\vspace{-1em}

CubeSats require robust OBDH solutions in harsh environments. The Demoiselle OBC, featuring a radiation-tolerant APSoC and layered FSW, supports reuse, in-orbit updates, and secure operations. To be validated through ITASAT2 and SelenITA, it ensures fault tolerance, flexibility, and compatibility with emerging technologies. This architecture establishes a foundation for long-lasting, scalable OBDH systems in future Brazilian CubeSat missions, ensuring long-term reliability and adaptability.

\end{abstract}

\section{INTRODUCTION}
    
CubeSats have emerged as essential platforms for space missions, offering compact design, lower costs, and capabilities once limited to larger spacecraft \cite{NAP2016achieving}. Yet, developing an on-board data handling (OBDH) subsystem that ensures fault tolerance (FT), adaptability, and reusability remains challenging due to the harsh environment and constraints of small satellites.

This paper presents a multimission OBDH solution for the ITASAT2 and SelenITA CubeSat missions, with the aim of supporting future Brazilian endeavors. Central to this approach is the Demoiselle on-board computer (OBC), inspired by Santos Dumont’s Demoiselle aircraft, which exemplified innovation, efficiency, and adaptability. The OBC and its flight software (FSW) emphasizes reconfigurability, radiation tolerance, and operation under extreme thermal conditions.

\begin{table} [!t]
    \centering
    \caption{Orbital parameters and simulation models}
    \resizebox{\columnwidth}{!}{%
    \begin{tabular}{c|c|c}
        \textbf{Parameters}  & \textbf{ITASAT2} & \textbf{SelenITA}\\
        \hline
        Central body & Earth & Moon \\
        Semi-major axis & 6778.14 km & 1888 km\\
        Eccentricity & 0º & 0º\\
        Inclination & 51.64º & 90º\\
        Lifetime & 2 years & 2 years\\
        Solar particle peak flux & CREME96 & CREME96\\
        Trapped radiation model & AP-8 Maximum & -\\
        Galactic cosmic radiation & ISO 15390 & ISO 15390\\
    \end{tabular}
    }
    \label{tab:mission_parameters}
\end{table}

\begin{figure}[t]
    \centering
    \includegraphics[width=\columnwidth]{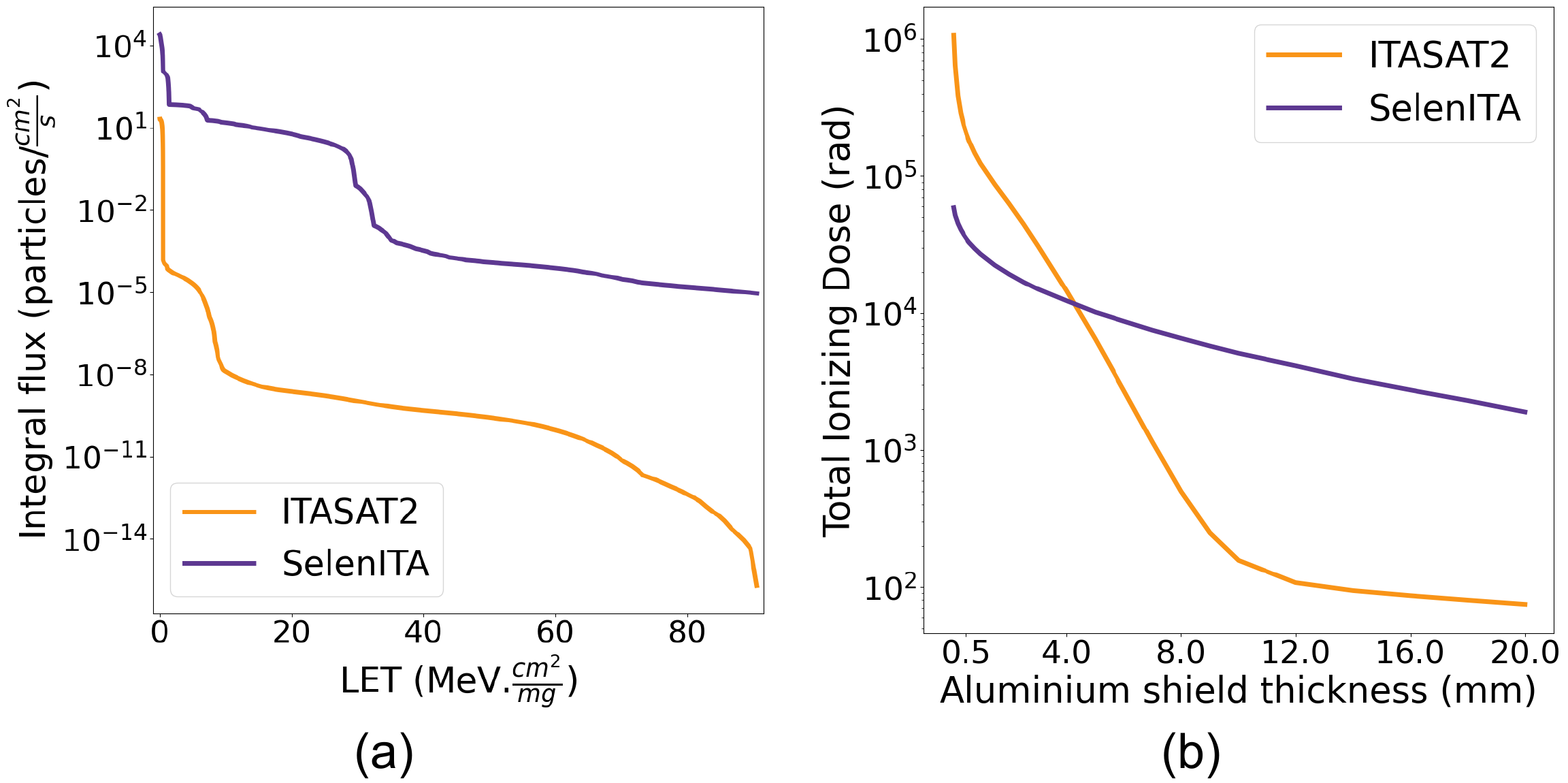}
    \caption{(a) Linear Energy Transfer (LET) spectra, and (b) Total Ionizing Dose (TID) as a function of Al shielding thickness.}
    \label{fig:mission_radiation}
\end{figure}

\section{SMALL SATELLITES AND MISSION CHALLENGES}

Small satellites, particularly CubeSats, have become a central element of contemporary space exploration and services \cite{NASA2023SOTA}. Originally introduced for academic and technology demonstration purposes \cite{puig2001cubesat}, CubeSats now support a wide range of applications in science, remote sensing, and communications. Missions typically follow three phases—conceptual design, detailed design and integration, and testing and deployment—ensuring systematic requirement flow-down \cite{shiotani2014end}.

The ITASAT2 \cite{loures2022itasat} and SelenITA \cite{matos2023engineering} programs, led by the ITA Space Center, exemplify the evolving capabilities and ambitions of CubeSat missions. ITASAT2 will deploy three 12U CubeSats in Low Earth Orbit (LEO) to study space weather and perform RF geolocation. SelenITA, a single 12U CubeSat, will orbit the Moon at Low Lunar Orbit (LLO) to conduct lunar geophysics and space weather investigations.

Operating in such demanding environments places stringent requirements on satellite subsystems, particularly OBDH. Spacecraft must withstand extreme thermal fluctuations, vacuum conditions, and intense radiation \cite{schwank1994}. Energetic particles induce displacement damage and total ionizing dose (TID), while single-event effects (SEE), like single-event upsets (SEU), can disrupt onboard electronics \cite{binder1975satellite,kastensmidt2006radiation}.

\begin{figure}[!b]
    \centering
    \includegraphics[width=0.9\columnwidth]{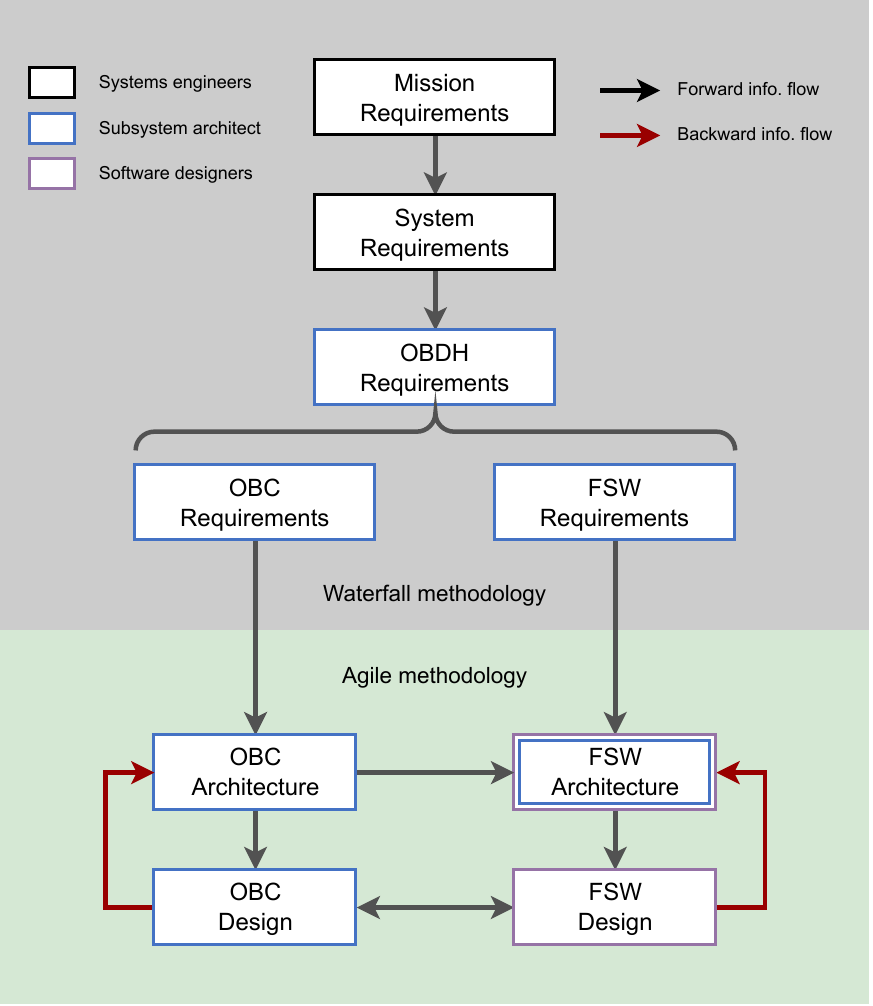}
    \caption{Requirements flow down for the OBDH subsystem.}
    \label{fig:req_flowdown}
\end{figure}

\begin{figure*}
    \centering
    \includegraphics[width=0.85\textwidth]{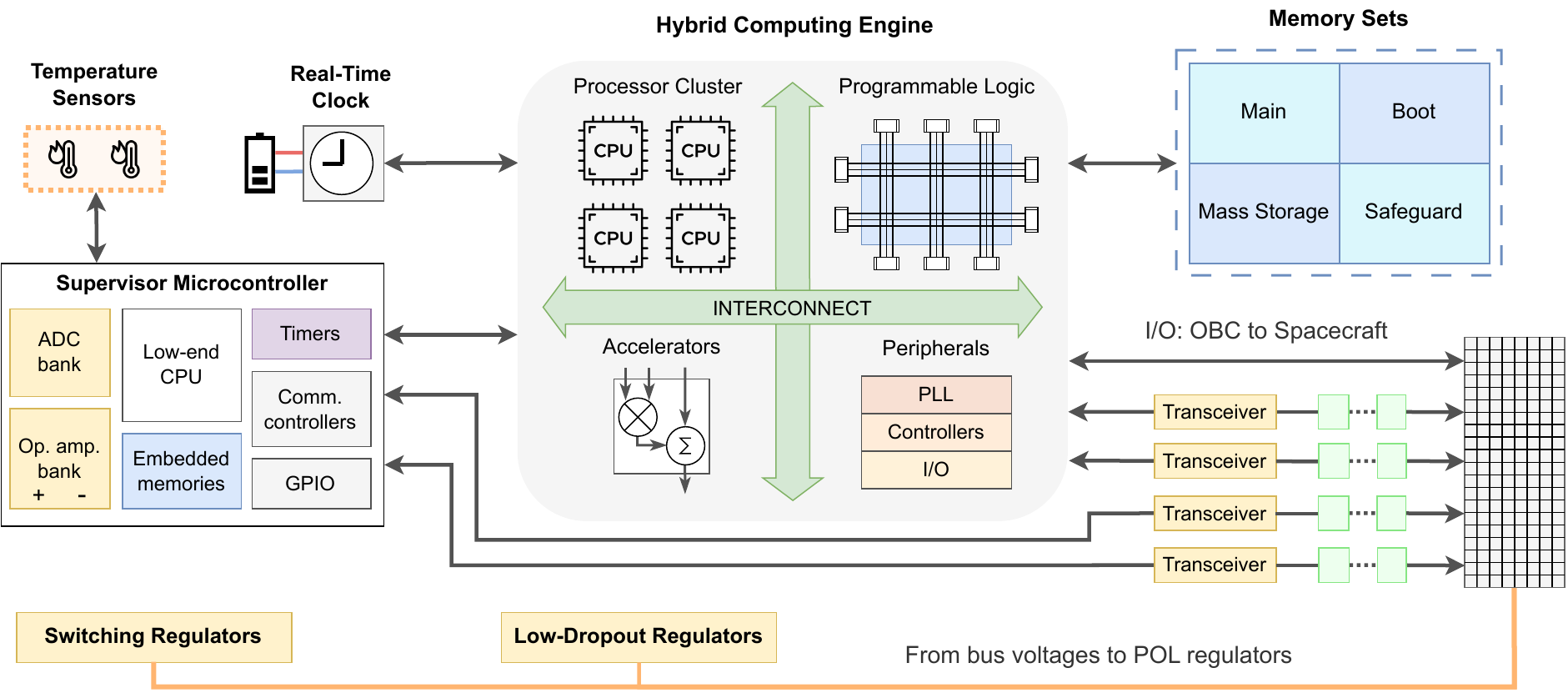}
    \caption{High-level architecture of the Demoiselle OBC.}
    \label{fig:demoiselle_architecture}
\end{figure*}

Figure~\ref{fig:req_flowdown} illustrates how OBDH requirements flow down from high-level mission objectives to subsystem specifications, guiding design choices and trade-offs. As shown in Figure~\ref{fig:demoiselle_architecture}, the Demoiselle OBC architecture incorporates strategies to address these challenges. Radiation simulations, generated using SPENVIS, highlight mission-specific conditions (Fig.~\ref{fig:mission_radiation}). In particular, the absence of a protective magnetosphere intensifies SelenITA’s radiation environment, while the South Atlantic Magnetic Anomaly introduces increased TID for ITASAT2.

In essence, the small size and cost-effectiveness of CubeSats enable more frequent and varied missions, but also demand robust OBDH solutions capable of coping with diverse radiation environments, thermal extremes, and operational constraints.

\section{ON-BOARD DATA HANDLING}

The OBDH subsystem manages payload data collection, timekeeping, subsystem communication, and the deployment of booms and antennas. The OBC, constrained by size, weight, power, and cost, serves as the spacecraft’s central hub.

As data demands grow, efficient bandwidth use and autonomy become crucial \cite{scarpa2021onthefly}, necessitating advanced processing and AI/ML capabilities \cite{leon2022towards}. Future missions, especially in challenging lunar environments, require high processing power, fault tolerance (FT), and reliability \cite{george2018onboard}.

The Demoiselle OBC and FSW development employs a structured requirement flow-down (Fig.\ref{fig:req_flowdown}) using a waterfall model for systematic traceability, while agile practices enable controlled iterative refinements. The backward arrows in Fig.\ref{fig:req_flowdown} illustrate how adjustments can flow upward, maintaining adaptability without overextending schedules, risks, or costs.

\section{Demoiselle OBC}

\begin{figure}[!b]
    \centering
    \includegraphics[width=0.8\columnwidth]{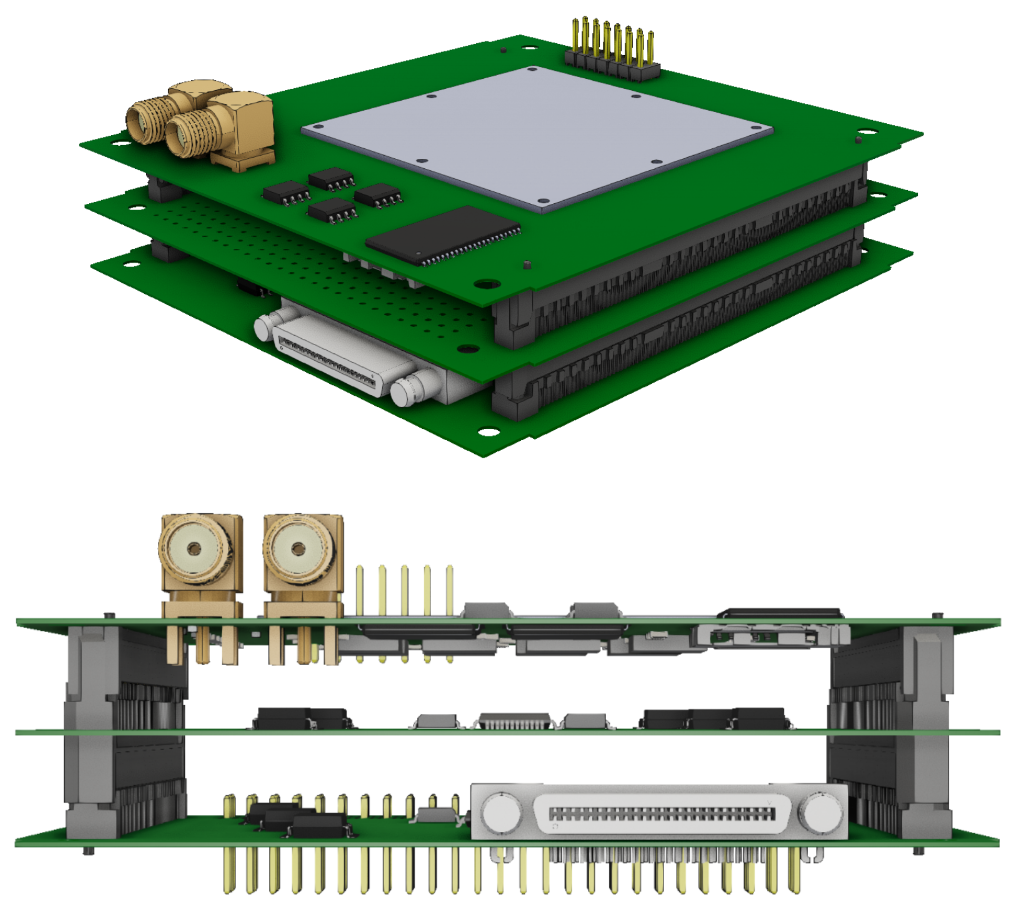}
    \caption{3D renders of the Demoiselle OBC, showing its three boards.}
    \label{fig:demoiselle_3d}
\end{figure}

\begin{figure*}
    \centering
    \includegraphics[width=\textwidth]{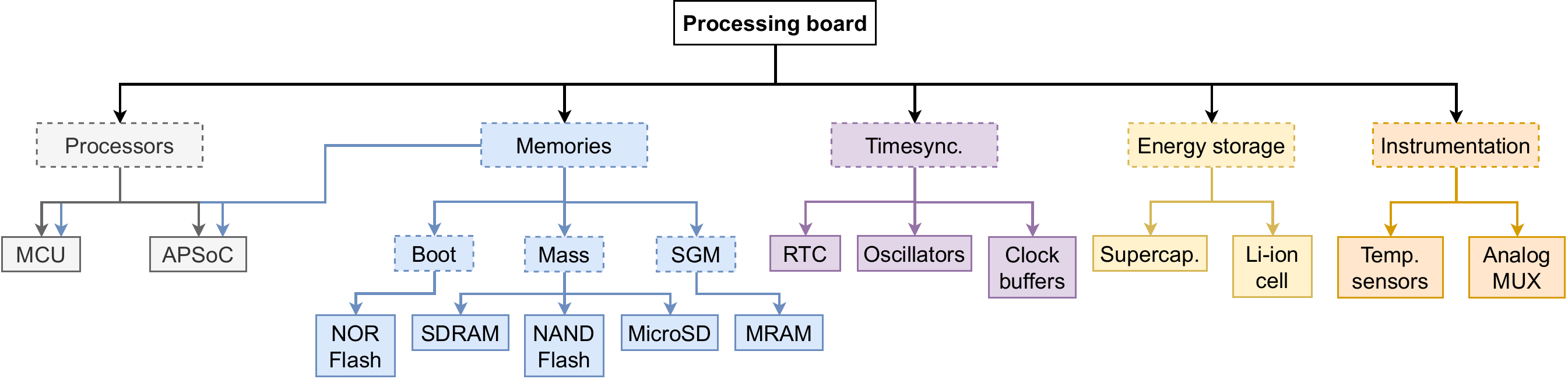}
    \caption{Component breakdown of the Demoiselle OBC processing board.}
    \label{fig:product_tree}
\end{figure*}

The Demoiselle OBC is designed for reuse across multiple missions, accommodating evolving requirements through a modular three-board configuration—processing, power, and interface boards—that ensures broad I/O compatibility (Figures~\ref{fig:demoiselle_architecture}, \ref{fig:demoiselle_3d}).
The component breakdown of the processor board is shown in Figure~\ref{fig:product_tree}.

To address the constraints of small satellites, the architecture minimizes reliance on radiation-hardened components and redundant boards, instead emphasizing careful component selection and system-level fault mitigation \cite{label1996single}.

The NanoXplore NG-Ultra, a radiation-hardened APSoC, has been selected as the hybrid computing engine (HCE) due to the advantages of APSoC over standalone MPSoC and FPGA solutions \cite{iturbe2016on}. This device is currently the only radiation-hardened APSoC available without export restrictions to Brazil \cite{costa2024heavy_mr}, creating potential supply risks for ITASAT2 and SelenITA. As a contingency, future iterations may employ standalone MPSoC and FPGA devices.

Previously integrated into Airbus’s OBC-Ultra for larger satellites \cite{danard2023obc}, NG-Ultra debuts here in a CubeSat context. Fabricated with 28 nm FD-SOI technology, it integrates four Cortex-R52 cores and an FPGA (537,000 LUT4) linked via AXI and APB buses, withstands up to 50 krad TID, exhibits no SEU susceptibility below 60 MeV·cm$^2$/mg, and incorporates multiple fault-tolerance mechanisms.

\subsection{Supervisory System}

Supervising complex processors with a simpler, dedicated unit is a well-established strategy in robust embedded system design \cite{oliveira2013reliability}. Here, a low-end, space-qualified microcontroller (MCU) serves as supervisor, handling sensor interfacing, external watchdog timers (WDT), and providing a redundant command path from ground stations, thereby enabling multiple fault handling strategies.

\begin{figure}[!b]
\centering
\includegraphics[width=\columnwidth]{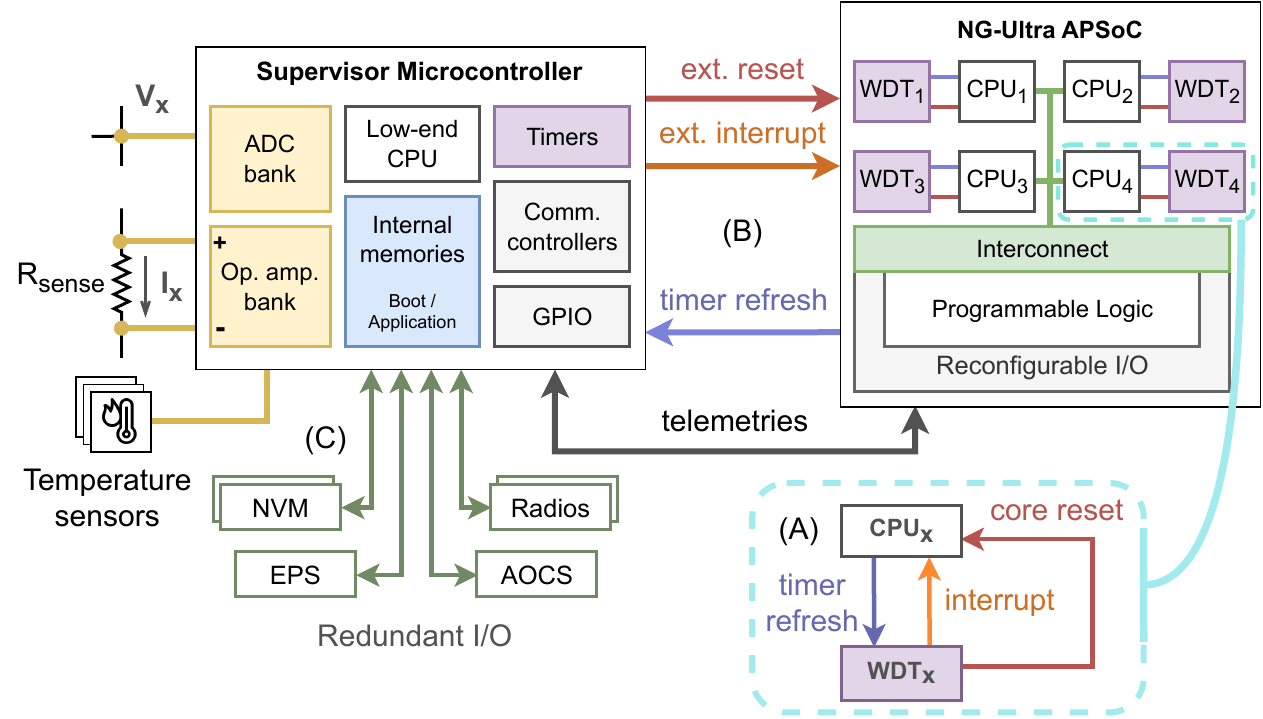}
\caption{Supervisor system for the Demoiselle OBC, providing system-level FT.}
\label{fig:demoiselle_supervisor}
\end{figure}

The WDT implementation employs three layers (Fig.~\ref{fig:demoiselle_supervisor}). Layer A utilizes the NG-Ultra's internal WDT for each CPU core, featuring two refresh thresholds: the first missed deadline triggers an interrupt, the second initiates a core reset. Layer B extends the period, relying on the supervisor MCU’s timers to externally reset the APSoC when not refreshed. Layer C, with the longest interval and most severe response, instructs the EPS to power-cycle the OBC if refresh signals remain absent.

\subsection{Memories}

The NG-Ultra memory controllers support DDR3/4 SDRAM with Reed-Solomon ECC and NOR Flash boot with triple modular redundancy (TMR). The eROM holds the Level 0 bootloader, while the SPI NOR Flash stores the Level 1 bootloader, FSW, and FPGA bitstream. Main memory is provided by NG-Ultra eRAM with single-error correction, double-error detection (SECDED), and scrubbing, while the TCM handles high-priority and real-time interrupts.

\begin{figure}[!t]
    \centering
    \includegraphics[width=0.9\columnwidth]{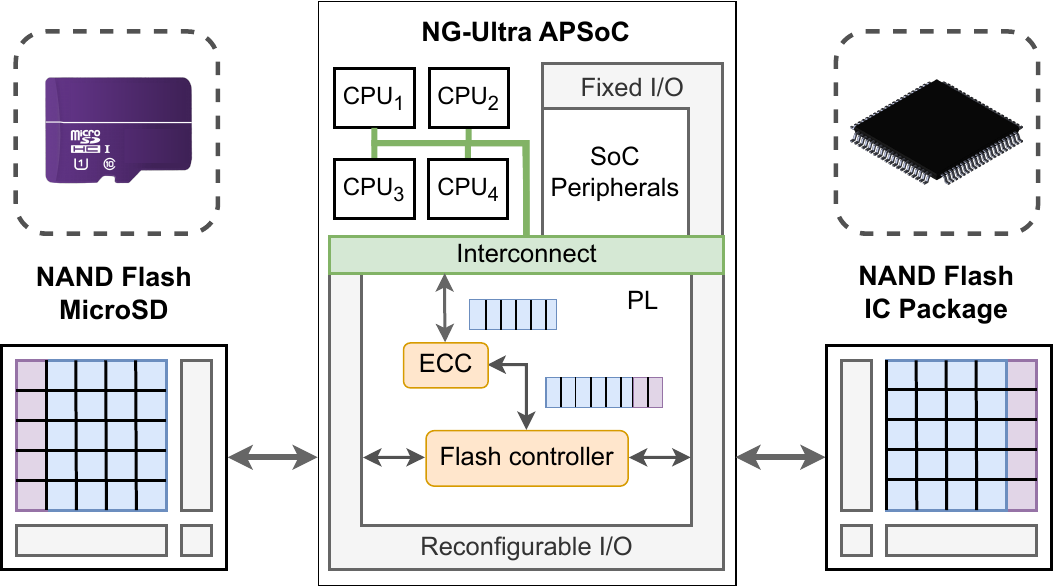}
    \caption{Mass memory of the Demoiselle OBC, featuring a NAND Flash memories.}
    \label{fig:demoiselle_mass_mem}
\end{figure}

Mass storage comprises DDR3/4 SDRAM, 3D NAND Flash, and a MicroSD card (Fig.~\ref{fig:demoiselle_mass_mem}). DDR3/4 SDRAM, protected by Reed-Solomon ECC, serves as a buffer. The 3D NAND Flash offers primary mass memory with CRC, while the MicroSD card provides redundancy and eases data handling during assembly, integration, testing, and in-orbit operations.

Modern NAND Flash in space must address TID susceptibility and SEU-related reliability trade-offs between SLC and MLC modes \cite{fabiano2013nand,wilcox2019radiation}, ensuring an optimal balance between density and robustness.

In LLO (SelenITA) and LEO (ITASAT2), GNSS signals may be unreliable or unavailable, prompting the OBC to autonomously generate pulse-per-second (PPS) signals to maintain timing accuracy. NG-Ultra periodically saves spacecraft state—telecommands, flags, counters—to a safeguard memory (SGM) with intra-device TMR and inter-device DMR. During power losses, a battery-backed RTC retains time. Upon reboot, the OBC restores state from the SGM and retrieves time from the RTC, ensuring seamless operational continuity.

\begin{figure}[!t]
    \centering
    \includegraphics[width=0.9\columnwidth]{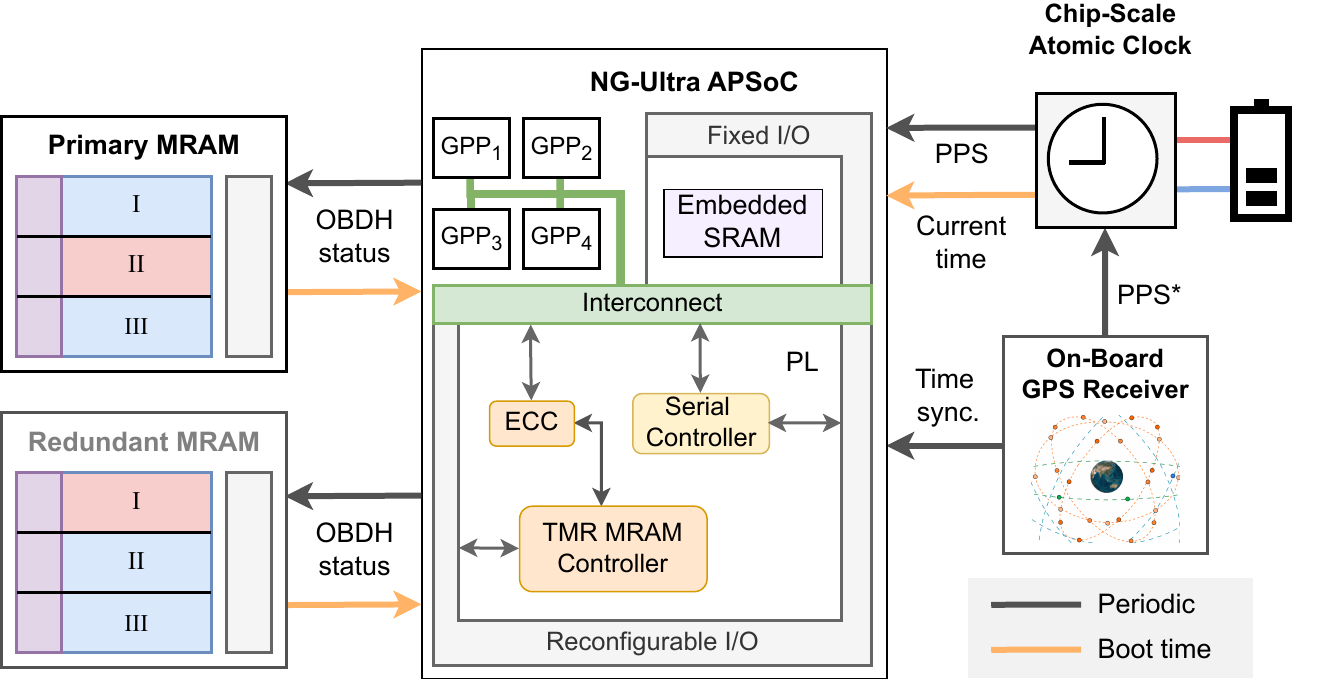}
    \caption{Reboot survival mechanism for the Demoiselle OBC, with SGM and RTC.}
    \label{fig:demoiselle_reboot_survival}
\end{figure}

As shown in Figure~\ref{fig:demoiselle_reboot_survival}, MTJ-MRAM is employed as SGM to provide high-density, durable non-volatile storage. Periodic state-saving to DMR MRAM and the use of RTC holdover support reliable reboots. Additional FT includes cache refreshes, conditioning of volatile and non-volatile memories, and detecting SEFIs in NAND Flash and MRAM. In such events, the HCE triggers resets or power cycles under MCU supervision.

\begin{figure*}
    \centering
    \includegraphics[width=\textwidth]{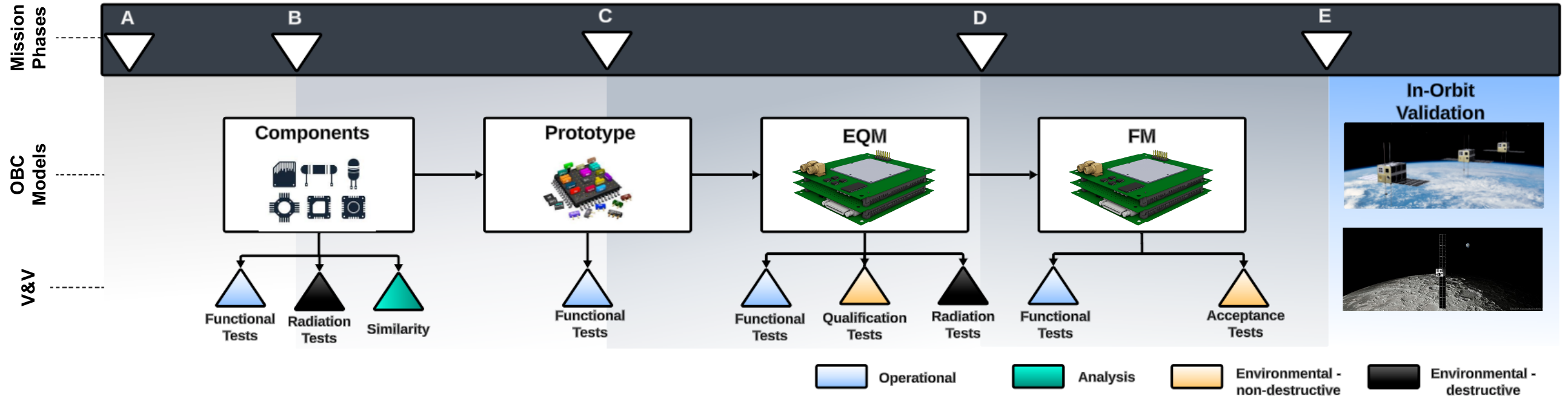}
    \caption{Lifecycle planning for the Demoiselle OBC, including the relationship between its models and mission phases, along with V\&V strategies for each model.}
    \label{fig:demoiselle_lifecycle}
\end{figure*}

\subsection{Power Conditioning}
    
The spacecraft bus provides limited voltages, which the OBC power board regulates and converts. The supervisor MCU, powered directly by the EPS, monitors OBC power lines and controls DC/DC converters to mitigate SEFI/SEL. Latch-up current limiters are not employed due to limited availability at CubeSat voltage levels, potential reboot loops \cite{label1996single}, and unsuitability for complex SoCs with workload-dependent power dissipation \cite{ferrandez2023case}. Future efforts will examine DC/DC derating and redundancy schemes to improve FT.

\subsection{Demoiselle OBC Lifecycle} \label{ssec:lifecycle}

The Demoiselle OBC follows NASA’s life cycle phases (A–E) \cite{hirshorn2017nasa} to achieve TRL 6 (Fig.~\ref{fig:demoiselle_lifecycle}). Each model—prototype, engineering/qualification model (EQM), and finally the flight model (FM)—undergoes verification and validation (V\&V) at increasing fidelity.

V\&V includes functional testing, radiation tolerance assessments \cite{sinclair2013radiation}, and environmental qualification (e.g., vibration, vacuum, thermal cycling). Acceptance tests ensure the system is ready for deployment. As ITASAT2 and SelenITA missions progress, lessons learned will guide refinements that further enhance robustness, flexibility, and performance

\section{MULTIMISSION FLIGHT SOFTWARE}

As space missions grow more complex and adopt common software platforms, ensuring reliable, secure, and adaptable flight software becomes critical. Evolving threats, supply chain vulnerabilities, and heightened autonomy requirements demand that FSW integrate safety, security, functionality, and performance standards. Guidelines such as MISRA-C, CERT-C, IEC 61508, and the NASA Software Safety Guidebook support robust development, while zero-trust principles help minimize attack surfaces.

The Demoiselle FSW emphasizes modularity and reusability, drawing on both in-house and open-source components. Over-the-air (OTA) updates enable in-orbit reconfiguration, from minor software routines to full RTOS and FPGA bitstream revisions. Scheduling updates to avoid high-radiation regions like the South Atlantic Magnetic Anomaly (SAMA) \cite{label1996single} enhances reliability.

\begin{figure}[!b]
    \centering
    \includegraphics[width=\columnwidth]{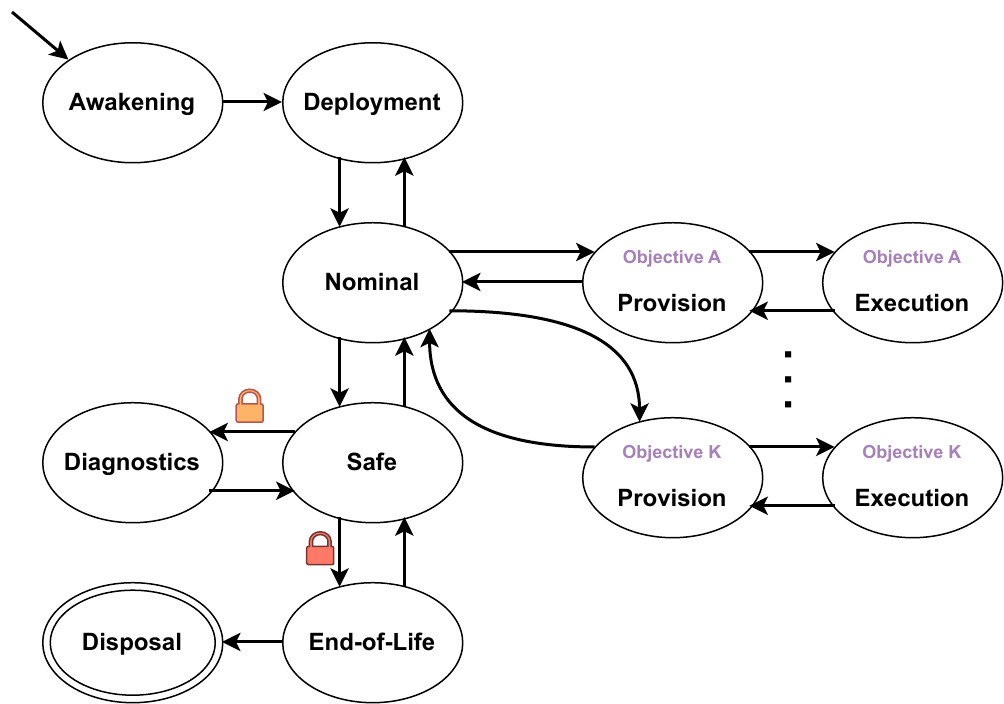}
    \caption{Operational modes of the FSW architecture, displaying the spacecraft lifecycle from LEOP to disposal.}
    \label{fig:fsw_op_modes}
\end{figure}

\begin{figure}[!b]
    \centering 
    \includegraphics[width=0.5\columnwidth]{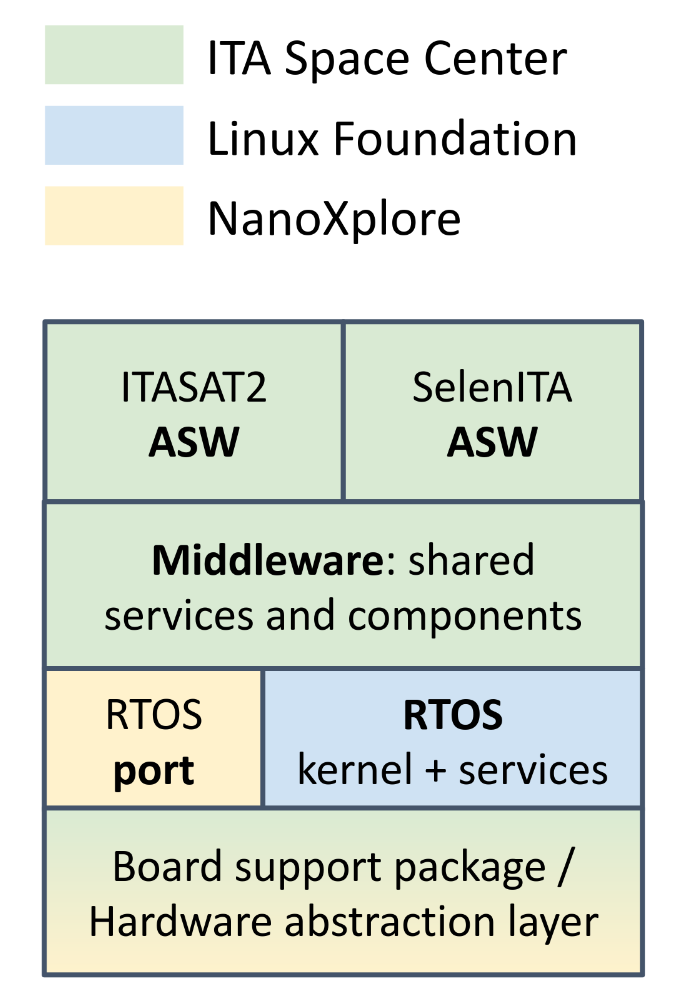}
    \caption{Layered architecture of the Demoiselle FSW stack.}
    \label{fig:fsw_layers}
\end{figure}

\subsection{Operational Modes}

Structured operational modes (Fig.~\ref{fig:fsw_op_modes})—awakening, deployment, nominal, safe, diagnostics, and end-of-life—optimize performance and resource allocation throughout the mission. Awakening/deployment prepare for operations, nominal handles routine tasks, safe limits functions after anomalies, diagnostics allows controlled analysis, and end-of-life oversees final disposal. This ensures alignment with mission phases while preserving safety, security, and adaptability

\subsection{Layered Architecture}

A layered architecture (Fig.~\ref{fig:fsw_layers}) supports long-term maintainability and scalability. The board support package (BSP) and hardware abstraction layer (HAL) ensure stable interfaces to the hardware. Above them, the RTOS kernel provides real-time performance, memory protection, secure multiprocessing, and hardware-enforced stack protection. The RTOS port layer adapts the kernel for mission-specific needs, while middleware standardizes common services to streamline integration across multiple missions.

At the top, the application software (ASW) layer incorporates mission-specific functionalities for platforms like ITASAT2 and SelenITA. This configuration enhances modularity, code reuse, and rapid adaptation to new payloads or mission objectives, all while adhering to functional safety principles.

\subsection{Security}

\begin{figure}[!t]
    \centering
    \includegraphics[width=0.9\columnwidth]{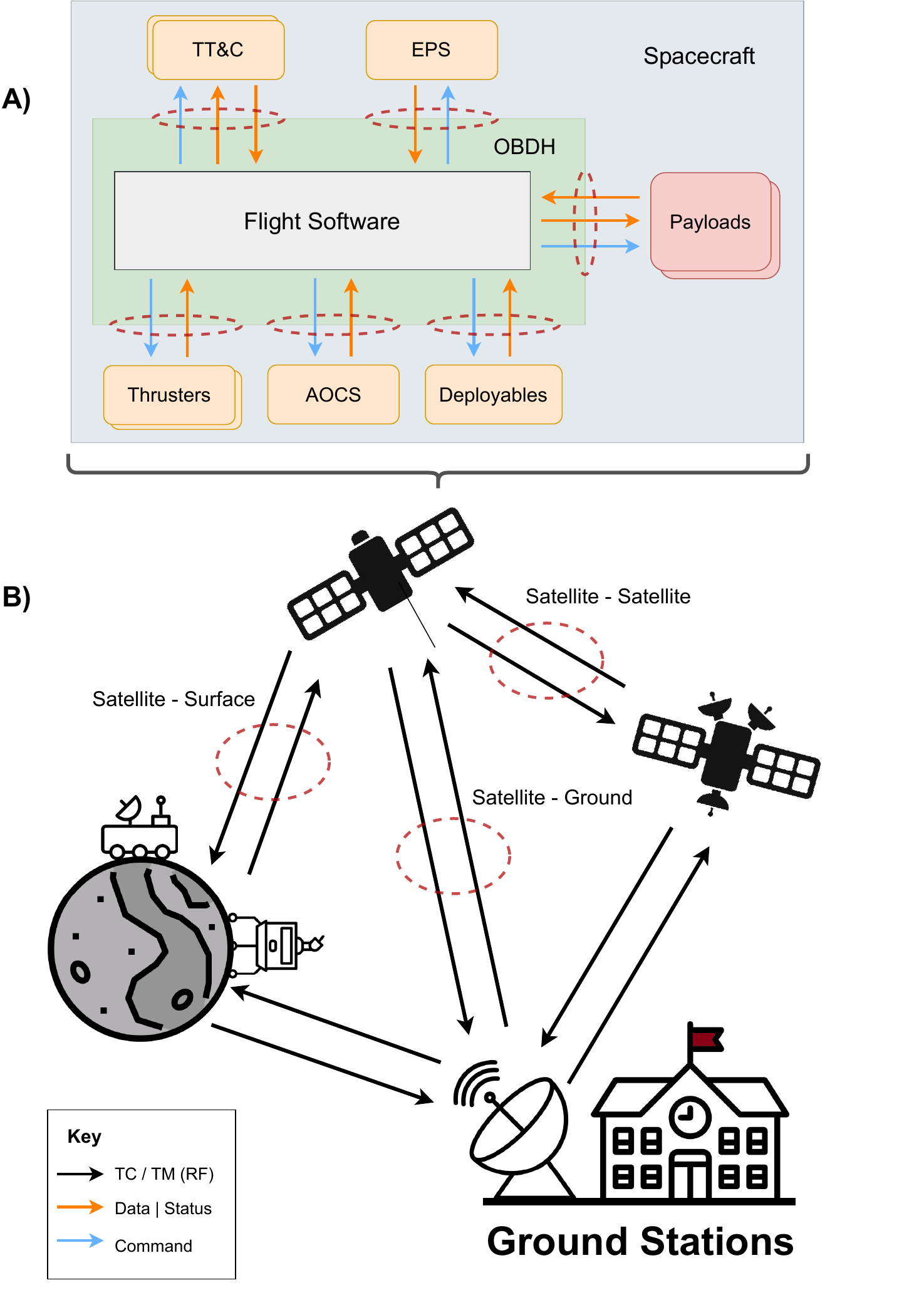}
    \caption{Context diagrams supporting threat modeling for space and link segments.} 
    \label{fig:context_diagram}
\end{figure}

Threat modeling, guided by context diagrams (Fig.~\ref{fig:context_diagram}), identifies vulnerabilities across space and link segments. Uniform security measures throughout Earth-orbit and lunar phases reduce exposure, while robust encryption, authenticated firmware updates, and strict access controls ensure data integrity and command authority.
Frameworks such as ESA’s SPACE-SHIELD and Aerospace’s SPARTA provide structured approaches to threat analysis, ensuring that uniform security measures

By integrating best practices for safety, security, and adaptability, the Demoiselle FSW mitigates evolving threats, supports continuous improvement, and meets the operational demands of challenging space environments.

\section{CONCLUSIONS}

This work presented the Demoiselle OBC and its FSW, designed for ITASAT2 and SelenITA with future reuse in mind. The hardware leverages a radiation-tolerant APSoC and modular architecture to achieve reliability and scalability. The FSW employs layered abstractions, operational modes, and OTA updates to maintain flexibility, security, and adherence to stringent safety standards.

As these missions advance, lessons learned will guide refinements in latch-up mitigation, redundancy strategies, and AI-driven on-board processing. The Demoiselle approach establishes a foundation for robust, reconfigurable OBDH subsystems that can adapt to evolving mission requirements and emerging technologies, ensuring longevity and operational success for future Brazilian CubeSat missions.

\printbibliography

\end{document}